# Introductory models of COVID-19 in the United States

**Peter Hugo Nelson**
Department of Physics, Fisk University, Nashville, TN 37208, USA
pete@circle4.com
*April 17, 2021. Revised August 28, 2021.*

## ABSTRACT

Students develop and test simple kinetic models of the spread of COVID-19 caused by the SARS-CoV-2 virus. Microsoft Excel is used as the modeling platform because it's non-threatening to students and because it's widely available. Students develop finite difference models and implement them in the cells of preformatted spreadsheets following a guided-inquiry pedagogy that introduces new model parameters in a scaffolded step-by-step manner. That approach allows students to investigate the implications of new model parameters in a systematic way. Students fit the resulting models to reported cases-per-day data for the United States using least-squares techniques with Excel's Solver. Using their own spreadsheets, students discover for themselves that the initial exponential growth of COVID-19 can be explained by a simplified unlimited growth model and by the SIR model. They also discover that the effects of social distancing can be modeled using a Gaussian transition function for the infection rate coefficient and that the summer surge was caused by prematurely relaxing social distancing and then reimposing stricter social distancing. Students then model the effect of vaccinations and validate the resulting SIRV model by showing that it successfully predicts the reported cases-per-day data from Thanksgiving through the holiday period up to February 14, 2021. The same SIRV model is then extended and successfully fits the fourth peak up to June 1, 2021, caused by further relaxation of social distancing measures. Finally, students extend the model up to the present day (August 27, 2021) and successfully account for the appearance of the delta variant of the SARS-CoV-2 virus. The fitted model also predicts that the delta-variant peak will be comparatively short, and the cases-per-day data should begin to fall off in early September 2021 – counter to current expectations. This case study makes an excellent capstone experience for students interested in scientific modeling.

# I. INTRODUCTION

When the COVID-19 pandemic reached the United States and classes were moved online, all of us had to reevaluate how we would teach and what content we would cover. I had been working on a long-term project to introduce molecular biophysics into the undergraduate curriculum using an active-learning approach – *Biophysics and Physiological Modeling* (BPM) (1). As an educator and modeler, I was curious as to whether the biophysical modeling techniques I had been developing for undergraduates using Excel (or compatible spreadsheet programs (2)) could be applied to modeling the spread of the COVID-19 disease caused by the novel SARS-CoV-2 virus. This article is an account of what I discovered with my students using the United States as a case study (3).

## A. Kinetic models

Kinetic models in biochemistry and biophysics apply to molecules being jiggled around by the molecules surrounding them. Similar models can – and have – been applied to a wide variety of other applications. Diffusion between two compartments can be modeled as a reversible first-order reaction (4). Drug elimination and radioactive decay are processes analogous to an irreversible first-order reaction (1). Population dynamics and epidemiological models can also be formulated using similar mathematical models. The approach we'll use is implemented as a simple finite difference (FD) model based on the Euler method, made famous by the 2016 movie *Hidden Figures*. These methods are ideal for introducing undergraduates to modeling kinetic processes because the computational steps of the FD model are represented by successive rows of the spreadsheet (1,4). Using Excel's Solver feature, the predictions of these FD models can be fitted to experimental data using least-squares techniques (1). As we'll discover, the same approach can also be used to model the spread of COVID-19 and compare the model predictions with reported data for confirmed cases of COVID-19 in the United States (3). It turns out that these simple FD models can do a surprisingly good job of modeling the spread of COVID-19 in the United States from February 2020 through August 2021.

## B. Learning objectives and pedagogical approach

The educational objectives are for students to learn how to:
1. Apply finite difference methods to introductory epidemiological models using the approach presented in ref. (1).
2. Apply systematic model development techniques to a complex real-world problem.
3. Use non-linear least-squares methods to test the predictions of the various numerical models by fitting them to published data for the United States.

The pedagogical approach is a guided-inquiry active-learning case study. Students begin by investigating the simplest possible "unlimited growth" epidemiological model. They then validate it by fitting it to published cases-per-day data for the United States as a whole. That model is then systematically modified to account for finite population size, recovery from COVID-19, changes in the infection rate coefficient due to changes in social distancing (and the delta variant), and finally, to account for vaccinations. The teaching materials use a scaffolded approach that focuses on the impact of each model parameter in a step-by-step manner (3).

The use of a systematic step-by-step approach is important for epidemiological models because they inherently produce exponential growth or decay in the infection rate. As a result, they are mathematically comparable to kinetic models of ion channel permeation that also predict exponential dependance (of electrical current on membrane voltage), and where it's been shown that the presence of too many parameters led to fitted parameters of questionable physical significance (5). Hence, in the modeling exercise presented here, students are only asked to add additional model parameters if the data call for them (Occam's Razor).

## II. FINITE DIFFERENCE MODELS AND FD DIAGRAMS

### A. Unlimited growth model

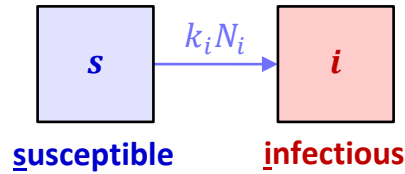

**Fig 1** FD diagram of the unlimited growth (UG) epidemiological model. The two boxes represent the two parts of the model population. Box $s$ represents people that are susceptible to the disease. Box $i$ represents people that are infectious.

In the simplest models of epidemiology, the population is split into two groups – susceptible and infectious (Fig 1). The simplest of those SI models is the unlimited growth (UG) model, in which the infection rate $R_i$ (arrow in Fig 1) is given by

$$R_i = k_i N_i \qquad (1)$$

where $k_i$ is the infection rate constant and $N_i$ is the number infectious. The idea behind the UG model and equation (1) is that an infectious person wanders randomly throughout the model population, just like a molecule in aqueous solution, infecting others with a rate characterized by an infection rate constant $k_i$, where $k_i = 0.25 \text{ d}^{-1}$ means that an infectious person infects a susceptible person every four days, on average, causing them to "jump" from box $s \to i$ when they become infectious (usually some days after contact with the infectious person).

The UG model of Fig 1 is unlimited because the model population is assumed to be infinite, and an infectious person stays infectious forever. Both of those assumptions are clearly incorrect, but they make for the simplest model. We'll discuss making the population finite, and modeling recovery in the next sections. However, the UG model gives students important insights into the initial spread of the virus – when the general population didn't know that SARS-CoV-2 was in their area.

Equation (1) is different from most first order reactions because the rate of jumps *into* box $i$ is proportional to the number $N_i$ already in box $i$. This can be contrasted with other processes like first-order drug elimination, where the rate of jumps *out of* the body (and into the bladder) is proportional to the number in the body (1,6).

According to the UG model of Fig 1, the finite difference equation for the small change $\delta N_i$ in the number infectious $N_i$ during a short time $\delta t$ (the timestep) is given by

$$\delta N_i = R_i \delta t \tag{2}$$

where $R_i$ is given by equation (1). Hence, the model can be implemented in a spreadsheet using the following condensed FD instructions (3)

$$t^{\text{new}} = t^{\text{old}} + \delta t \tag{3}$$

$$R_i^{\text{new}} = k_i N_i^{\text{old}} \tag{4}$$

and

$$N_i^{\text{new}} = N_i^{\text{old}} + R_i^{\text{new}} * \delta t \tag{5}$$

where the superscript $^{\text{old}}$ refers to the previous row in the spreadsheet (previous computational step) and $^{\text{new}}$ refers to the current row of the spreadsheet (current computational step). Hence, $N_i^{\text{old}}$ is the old number infectious (previous row at time $t^{\text{old}}$) and $N_i^{\text{new}}$ is the new number infectious (current row at time $t^{\text{new}} = t^{\text{old}} + \delta t$) – see the glossary of symbols in the appendix. The first activity asks students to write out a complete FD algorithm based on equations (3), (4) and (5) to calculate the infection rate $R_i(t)$ and number infectious $N_i(t)$. Students implement the algorithm in the rows of a preformatted spreadsheet and, by plotting the model predictions on semi-log graphs, they discover that the UG model predicts an exponential growth in both the number infectious $N_i$ and the infection rate $R_i$ (3,8).

In a "show that" exercise, students solve the elementary differential equation implied by equations (1) and (2) to give the following analytical solution

$$N_i = N_0 e^{k_i t} \tag{6}$$

which predicts an exponential growth in the number infectious from an initial number infectious $N_0$ (at $t = 0$). Substituting equation (6) into equation (1) yields

$$R_i = k_i N_0 e^{k_i t} \tag{7}$$

for the infection rate $R_i(t)$. Equation (7) is important because $R_i$ corresponds to the publicly available number of new confirmed COVID-19 cases reported per day. Because both $N_i$ and $R_i$ are exponential functions of time, students are able to show that the exponential growth can be characterized by a doubling time

$$t_d = \frac{\ln 2}{k_i} \tag{8}$$

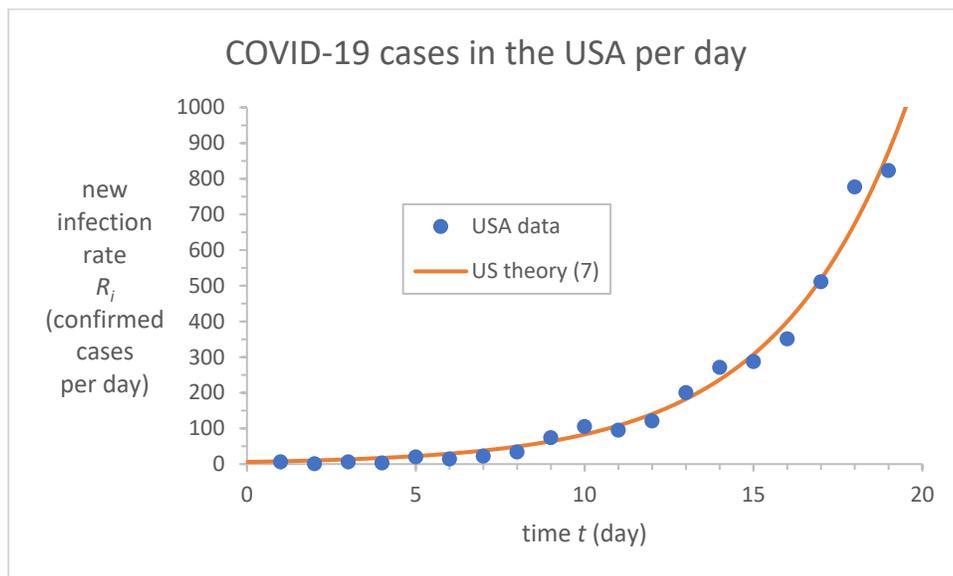

**Fig 2** Excel chart comparing the exponential growth model of equation (7) for $R_i(t)$ with reported data for the USA in the 19 days after February 26, 2020. The solid line is a least-squares fit to the USA data. The fitted model parameters are $N_0 = 23$ and $k_i = 0.26 \text{ d}^{-1}$ (data source ECDC (9)).

Students then compare the predictions of equation (7) with published cases-per-day data using a preformatted spreadsheet: first using Excel's "exponential trendline"; and then using least-squares techniques implemented using Excel's Solver. Fig 2 shows the resulting least-squares fit to equation (7) with data reported for the US during the 19 days after February 26, 2020. As students discover, the model does a surprisingly good job of explaining the reported data, validating the UG model's prediction of exponential growth for the initial uncontrolled spread of the contagion (3).

## B. Finite population model

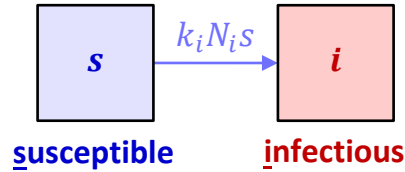

**Fig 3** FD diagram of a two-box epidemiological model exhibiting limited growth. The two boxes in this finite population (FP) model represent the two parts of the population that can be affected by the disease. Box $s$ represents the portion of the population susceptible to the disease. Box $i$ represents the portion infectious. Lowercase $s$ is the fraction of the population that are still susceptible to infection.

FD diagram Fig 3 shows a simple modification of the UG model that accounts for the finite size of the population. In this finite population (FP) model, the infection rate is given by

$$R_i = k_i N_i s \tag{9}$$

where $s$ is the susceptible fraction of the population that's *defined* by

$$s \equiv \frac{N_s}{N} \tag{10}$$

where $N_s$ is the number susceptible and $N$ (with no subscript) is the total number of people in the model population, where

$$N = N_s + N_i \tag{11}$$

Hence, we're assuming that the model population size doesn't change during the modeling time (no births or deaths). As a result, we can update $N_s$ using the instruction

$$N_s^{\text{new}} = N - N_i^{\text{new}} \tag{12}$$

The idea behind equation (9) is that people behave like molecules in solution. They randomly bump into each other at a constant rate, on average. Infections occur with a fixed probability when people get close together. If we assume that encounters occur at a constant rate, then the probability that an infectious person interacts with a susceptible person (as opposed to another infectious person) is simply $s$, the fraction of the population that's susceptible. In other words, $s$ is the fraction of people an infectious person encounters that are still susceptible to the virus. These simple assumptions are easy to understand, but it's important to remember that people aren't molecules.

By substituting the definition (10) of $s$ into equation (9) and solving equation (11) for $N_s$, students show that the finite population model can be calculated using

$$R_i^{\text{new}} = k_i * N_i^{\text{old}} * N_s^{\text{old}}/N \qquad (13)$$

and instructions (5) and (12).

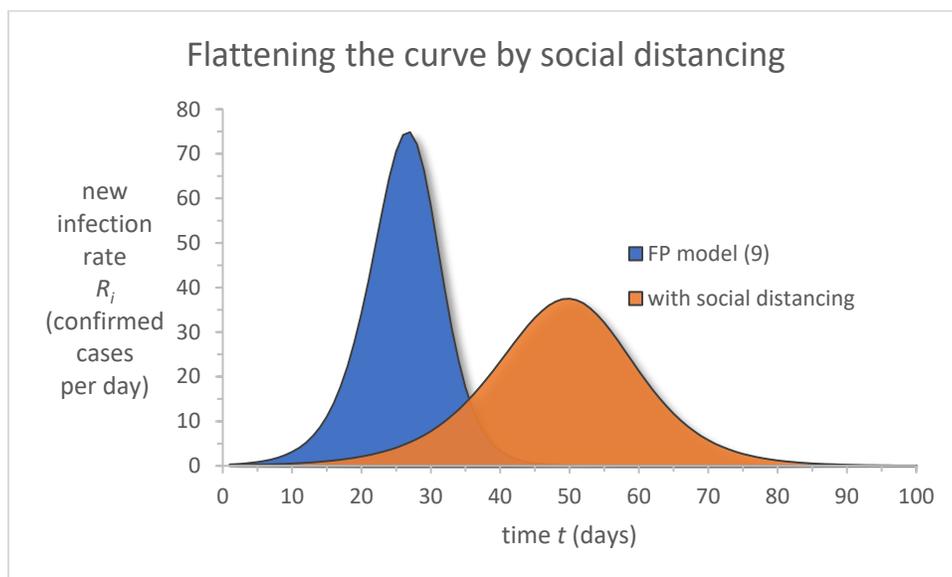

**Fig 4** Excel area chart showing the predictions of the finite population (FP) model for $R_i(t)$ (9) for a model population of $N = 1000$, an infection rate constant of $k_i = 0.3 \text{ d}^{-1}$, an initial number infectious of $N_0 = 1$ and a timestep of $\delta t = 1$ d. The "with social distancing" curve shows the effect of reducing the infection rate constant by a factor of 2 on day 0 to $k_i = 0.15 \text{ d}^{-1}$ by implementing social distancing and mask wearing.

Students write out a complete algorithm for the FP model and implement it in a preformatted spreadsheet. They are then able to investigate how social distancing can "flatten the curve" by halving the infection rate constant from $k_i = 0.3 \text{ d}^{-1}$ to $k_i = 0.15 \text{ d}^{-1}$ as shown in Fig 4. The effect of having a finite population is that the infection rate no longer increases exponentially without limit and there is a peak in the $R_i(t)$ curve that can be flattened by social distancing. Interestingly, the $N_i(t)$ curve (not shown) exhibits the classic logistic growth first reported by Verhulst and later by McKendrick (10). According to the FP model, social distancing merely delays the inevitable. Eventually, everyone in the model population becomes infectious despite the reduced infection rate constant. As we'll discuss in the next section, the SIR model is qualitatively different, because social distancing can reduce the ultimate number infected, and it can even prevent the outbreak from occurring at all.

## C. SIR model

The main problem with the finite population model is that people don't recover – ever. Clearly, that's not realistic. People do recover from COVID-19 and hence stop being infectious after a period of time. Fig 5 shows an FD diagram for the SIR epidemiological model. Named after the letters used for the three boxes, it's an important base model in epidemiology developed by Kermack and McKendrick in 1927 (11).

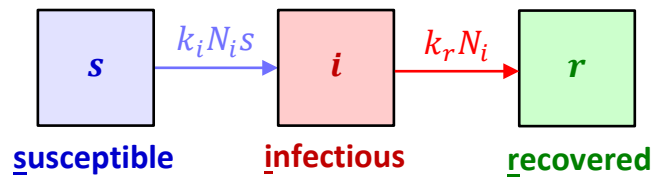

**Fig 5** FD diagram of the SIR epidemiological model. The three boxes represent the three parts of the model population that can be affected by the disease. Box $s$ represents the portion that's susceptible to the disease. Box $i$ represents the portion infectious. Box $r$ represents the portion that's recovered from the infection (or died). Sometimes this box is labeled removed – as in removed from consideration.

The three boxes in Fig 5 represent the possible states of people in the model population. In the SIR model, $N_s$, $s$ and $N_i$ have the same meaning as the FP model. The new state variable is $N_r$, which represents the number recovered. It's the number of individuals in the model population that have been infected – but have now recovered and are no longer infectious and are further assumed to be immune to the disease forever. The symbol $N_r$ more correctly stands for the number removed from the susceptible or infectious boxes. In addition to recovering, individuals can be removed from the number infectious by being isolated or quarantined from the susceptible portion of the population and they are also removed by death. All those individuals are represented by box $r$. We also have a relationship with the total number $N$ in the model population and it spells out the initials of the SIR model in the subscripts of the bookkeeping equation

$$N = N_s + N_i + N_r \qquad (14)$$

Now that we've discussed the three boxes, let's talk about the arrows between boxes in Fig 5. The first arrow from box $s \to i$ represents the rate of infection $R_i = k_i N_i s$ (9) – the exact same equation that we used for the finite population model of Fig 3. The second arrow from box $i \to r$ represents the rate of recovery. That recovery rate is given by

$$R_r = k_r N_i \qquad (15)$$

where $k_r$ is the recovery rate constant and the mean residence time in box $i$ (1) is predicted to be

$$\tau_i = \frac{1}{k_r} \qquad (16)$$

We'll call $\tau_i$ the mean infectious time. It can be approximated by a quantity that can be measured clinically – the mean "recovery time".

Using the information above, students write out the following FD instructions for the SIR model:

$$R_i^{new} = k_i * N_i^{old} * N_s^{old}/N \qquad (13)$$

$$R_r^{new} = k_r * N_i^{old} \qquad (17)$$

$$N_i^{new} = N_i^{old} + (R_i^{new} - R_r^{new}) * \delta t \qquad (18)$$

$$N_r^{new} = N_r^{old} + R_r^{new} * \delta t \qquad (19)$$

and

$$N_s^{new} = N - N_i^{new} - N_r^{new} \qquad (20)$$

They then write out a complete algorithm for the SIR model and implement it in a preformatted spreadsheet and then they answer a series of questions comparing the properties of the SIR model with the previous models.

Subsequently, students investigate the effects of social distancing by reducing the infection rate constant $k_i$. Fig 6 shows the kind of graphical information students observe in their spreadsheets. As shown in Fig 6, they discover that not everyone in the model population needs to be infected by the end of the pandemic if social distancing is implemented and maintained until the pandemic has subsided, i.e., the number susceptible $N_s$ doesn't reach zero at the end of the pandemic, meaning that some susceptible individuals were never infected (see week 50 in Fig 6(a)).

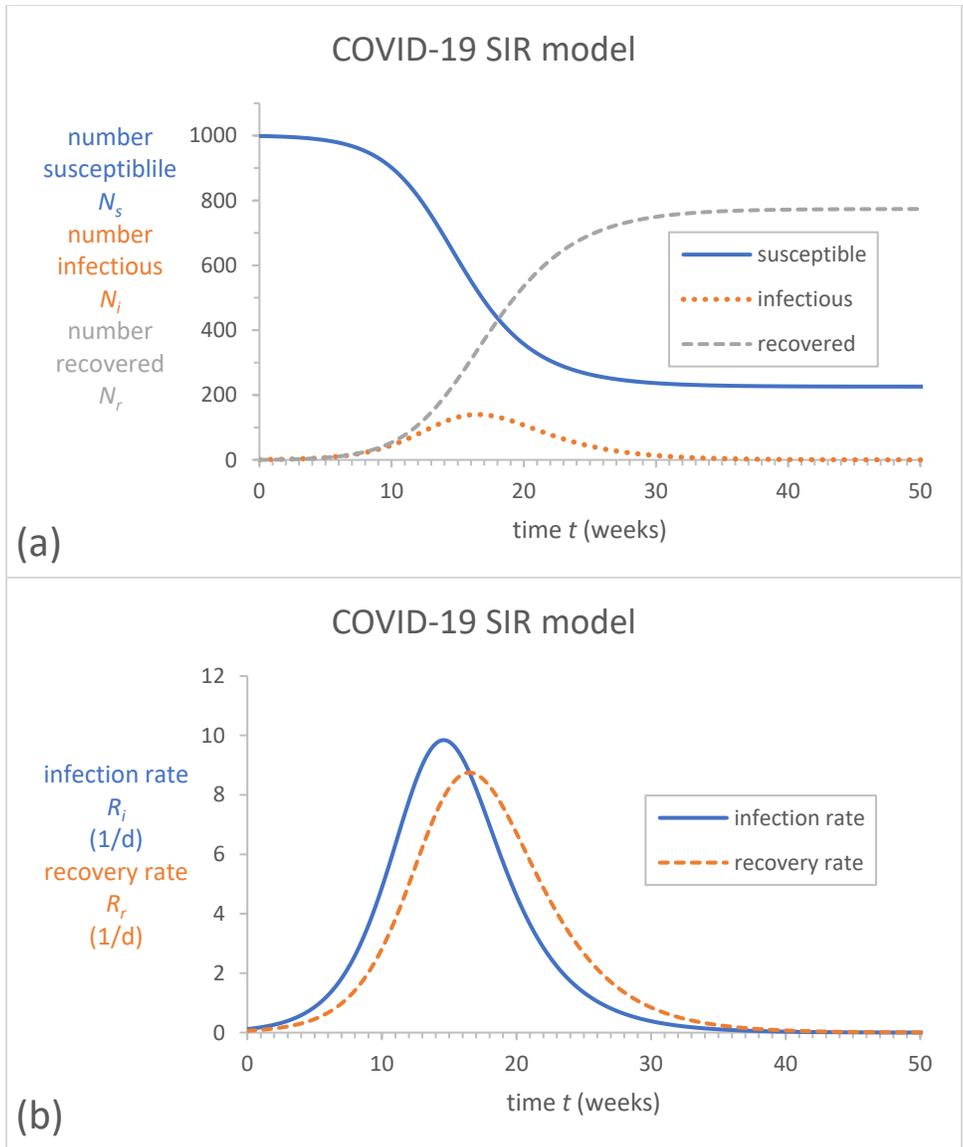

**Fig 6** Excel charts showing the predictions of the SIR model for a model population of $N = 1000$, an infection rate constant of $k_i = 0.12 \text{ d}^{-1}$, a mean infectious time of $\tau_i = 16$ d, an initial number infectious of $N_0 = 1$ and a timestep of $\delta t = 0.01$ d. Chart **(a)** shows the numbers in the three boxes $s$, $i$ and $r$ of the SIR model. Chart **(b)** shows the exponential dragon predicted for the infection rate $R_i$ (solid blue line) and the recovery rate $R_r$. **Note:** because of equation (15), the recovery rate $R_r$ is directly proportional to the number infectious $N_i$.

A universal feature of the SIR model is that it predicts an "exponential dragon" (12) in the infection rate $R_i(t)$ (Fig 6(b) and Fig.12.15 of ref. (3)), whose duration is determined by the value of the infection rate constant $k_i$. The peak of the exponential dragon exhibits a characteristic inverted vee shape on a semi-log plot (3).

In a guided-inquiry exercise, students investigate the effect of the model population size $N$ on the predictions of the model. They discover that model populations of all sizes in the SIR model behave in a similar manner and produce the same shape curves for $N_s$, $N_i$, $N_r$, $R_i$, and $R_r$, etc. – independent of the size of the model population. In a later "show that" problem, they discover that the SIR model can be reformulated in terms of fractional variables $s$, $i$ and $r$, where the fraction susceptible $s$ is defined by equation (10), the fraction infectious $i$ is defined by

$$i \equiv \frac{N_i}{N} \tag{21}$$

and the fraction recovered $r$ is defined by

$$r \equiv \frac{N_r}{N} \tag{22}$$

The fact that the SIR model predicts the same behavior independent of model population size is an important property of the SIR model.

## D. Herd immunity

After the peak in $N_i(t)$, the SIR model predicts that the number infectious $N_i$ will steadily decline because the rate of infection $R_i$ is less than the rate of recovery $R_r$. This can be related to the epidemiological concept of herd immunity (13). One way to see how the two concepts are related is to consider the quantity $1 - s_p$, which is the cumulative fraction that have been infected at the time $t_p$ of the peak in $N_i(t)$. We can then define $i_p$ as the fraction infectious and $r_p$ as the fraction recovered at time $t_p$, respectively. Hence, from the bookkeeping equation (14) we have

$$s_p + i_p + r_p = 1 \tag{23}$$

so that $1 - s_p = i_p + r_p$. Hence, the quantity $1 - s_p$ is the sum $i_p + r_p$, which is the cumulative total number of people that have been infected at time $t_p$.

Our SIR model assumes that individuals who've been infected can't be infected again. Hence, anyone who's already been infected is permanently immune in our SIR model. As a result, the fraction of the model population that are immune at any time can be written as

$$h = i + r = 1 - s \tag{24}$$

where $h$ is the fraction immune, or the immune fraction of the model population. Once the immune fraction $h$ reaches

$$h_p = 1 - s_p \tag{25}$$

the recovery rate $R_r$ is larger than the infection rate $R_i$ and the model predicts that the disease will be in decline and eventually die out. The fraction $h_p$ is the herd immunity threshold. If the fraction immune is greater than or equal to $h_p$, i.e., if

$$h \geq h_p \tag{26}$$

then the disease will be in decline rather than growing (as indicated by whether $N_i(t)$ decreases or increases, respectively).

### E. Finding the peak in the curve and $\mathcal{R}_0$

In a guided-inquiry exercise, students are asked to consider why the infection rate curve $R_i(t)$ always cuts through the peak in the recovery rate curve $R_r(t)$ – see for example Fig 6(b). Students discover that the peak in $N_i(t)$ and $R_r(t)$ occurs when

$$R_i = R_r \tag{27}$$

By substituting equations (9) and (15) into equation (27), students show that the value of the fraction susceptible $s$ at the peak in $N_i$ is given by

$$s_p = \frac{k_r}{k_i} = \frac{1}{\mathcal{R}_0} \tag{28}$$

where $\mathcal{R}_0$ is the basic reproduction number that's given by

$$\mathcal{R}_0 \equiv k_i \tau_i = \frac{k_i}{k_r} = \frac{1}{s_p} \tag{29}$$

The basic reproduction number $\mathcal{R}_0$ was made famous in the 2011 movie *Contagion*, and it's arguably the most important widely discussed parameter of epidemiological models (14). The first part of equation (29), $\mathcal{R}_0 \equiv k_i \tau_i$, defines $\mathcal{R}_0$ as the number infected by a single individual at the beginning of the outbreak when $s \approx 1$. $k_i$ is the average number of people infected by a single infectious individual per day and $\tau_i$ is the mean number of days that they're infectious.

Because $\mathcal{R}_0 = 1/s_p$, we can write the herd immunity threshold $h_p$ (25) in terms of the basic reproduction number $\mathcal{R}_0$

$$h_p = 1 - \frac{1}{\mathcal{R}_0} \tag{30}$$

The best-fit value of $\mathcal{R}_0$ students obtain by fitting the USA data with $\tau_i = 8$ d is $\mathcal{R}_0 \approx 4.1$ so that the herd immunity threshold for the original variants of COVID-19 is $h_p \approx 0.76$ or about 76% – in the absence of any social distancing measures. Recall that in the SIR model, the value of the infection rate constant $k_i$ depends on the level of social distancing. Hence, whenever the infection rate $R_i(t)$ is decreasing we have technically passed the herd immunity threshold for the current value of $k_i$.

## III. GAUSSIAN TRANSITION FUNCTIONS

The ultimate goal of this section is to model transitions between different levels of social distancing in a straightforward manner by making the infection rate coefficient a function of time $k_i(t)$ in the SIR model (15). The title of this section is a spoiler because it really wasn't obvious – at least to me – that Gaussian transition functions between the different epochs in the pandemic were a good way to go, even though, in hindsight it seems rather obvious.

### A. Initial transition to social distancing

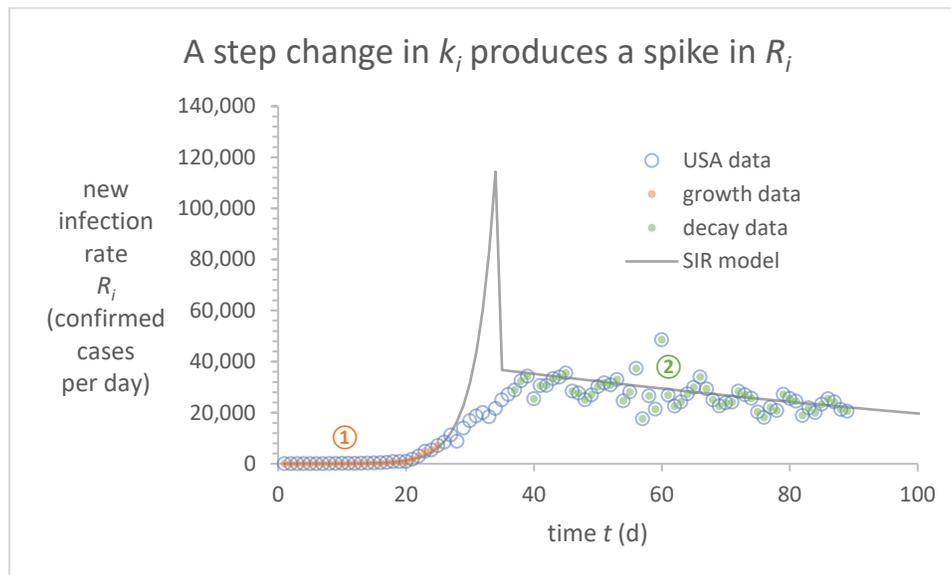

**Fig 7** Excel chart showing the fitted SIR model with a step change in the infection rate coefficient from $k_i = k_1 = 0.505$ d$^{-1}$ to $k_i = k_2 = 0.118$ d$^{-1}$. $k_1$ is calculated from the initial epoch ① of exponential growth (orange dots) and $k_2$ is the value students calculated for the epoch ② of social distancing (green dots). The only adjustable parameter in the fit is the transition time $t_{12} = 35$ d between $k_1$ and $k_2$. The other parameters in the model are $N = 8.25 \times 10^7$, $\delta t = 1$ d, $\tau_i = 8$ d, and $N_0 = 5.54$ (3). As shown, the fit produces a transition spike (grey line) that doesn't match the reported transition data (data source ECDC (9)).

When students get to this section, they already know that the SIR model can successfully model the pandemic in the US during the initial period of exponential growth (epoch ① in Fig 7) using an infection rate constant of $k_i = k_1 = 0.505$ d$^{-1}$. In a guided-inquiry activity, students discover that the SIR model can successfully model the exponential decay during the epoch ② of social distancing up to Memorial Day, May 25, 2020. The fit is excellent, but the initial conditions for the fit are arbitrary and meaningless. My first attempt at modeling the transition using a step change in the infection rate coefficient $k_i$ was an abject failure. As shown in Fig 7, the model can successfully model both the epoch ① of exponential growth and the epoch ② of social distancing – but the step-change in $k_i$ between epochs ① and ② produced a spike in the fitted model of $R_i(t)$ that's clearly inconsistent with the reported data during the transition period.

    As the title of this section states, the solution to this conundrum was to change the transition function for $k_i(t)$ from a step change to a Gaussian transition function. Once again, it turns out that Excel is a convenient platform for modeling COVID-19 because it includes a function NORM.DIST that implements the Gaussian distribution, both as a probability density function and as a cumulative probability. The latter is what we need for our Gaussian transition function. The justification for a Gaussian transition function is that not all states, communities or individuals took up social distancing etc. at the same time (or to the same extent). The simplest assumption is that those transition times are normally distributed and hence can be represented by a Gaussian transition function (16). From a modeling perspective, a Gaussian transition function is appealing because it introduces only one additional parameter for the standard deviation of the Gaussian that accounts for the statistical spread in the times when individual people changed their level of social distancing. The implementation that students use in Excel has two parameters: $t_{12}$ is the mean transition time between epochs ① and ②; and $\sigma_{12}$ is the standard deviation of the distribution of transition times between epochs ① and ②. The Excel function for calculating the cumulative probability $F_{12}(t)$ of the normal (Gaussian) distribution for the transition times between epochs ① and ② is

$$F_{12}^{\text{new}} = \text{NORM.DIST}(t^{\text{new}}, t_{12}, \sigma_{12}, \text{TRUE}) \qquad (31)$$

where $F_{12}^{\text{new}}$ is the value of the Gaussian cumulative probability at time $t^{\text{new}}$ and TRUE is the value of the "cumulative" parameter of the NORM.DIST Excel function (3,17). The corresponding probability density $p_{12}(t)$ can be calculated using cumulative = FALSE, i.e.

$$p_{12}^{\text{new}} = \text{NORM.DIST}(t^{\text{new}}, t_{12}, \sigma_{12}, \text{FALSE}) \qquad (32)$$

The time dependent infection rate coefficient $k_i(t)$ can then be calculated using

$$k_i^{\text{new}} = k_1 + F_{12}^{\text{new}} * (k_2 - k_1) \qquad (33)$$

Students implement the model in a preformatted spreadsheet and fit the model to the published data using least-squares techniques and Excel's Solver (Fig 8) (3,17).

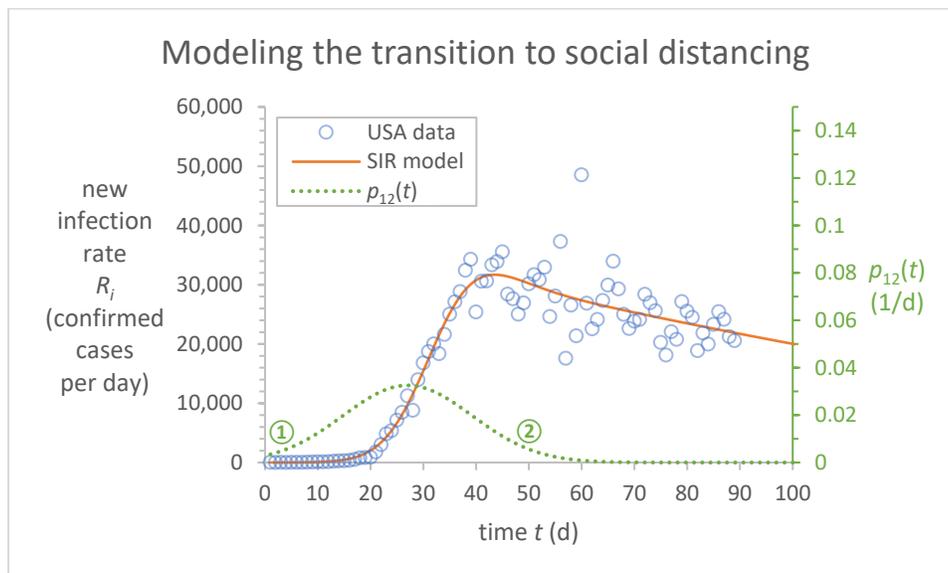

**Fig 8** Excel chart showing the SIR model (solid orange line) fitted to the USA data (blue circles). SIR model parameters $k_1 = 0.60$ d$^{-1}$, $k_2 = 0.12$ d$^{-1}$, $t_{12} = 27$ d and $\sigma_{12} = 12.3$ d were fit simultaneously using the least squares (LS) method. The remaining SIR model parameters were set to $N = 8.25 \times 10^7$, $\delta t = 1$ d, $\tau_i = 8$ d and $N_0 = 5$. The transition from $k_1 \rightarrow k_2$ is modeled using a Gaussian (normal) distribution with mean $t_{12}$ and standard deviation $\sigma_{12}$. The dotted line shows $p_{12}(t)$, the probability density function (32) of the Gaussian transition function $F_{12}(t)$ (31) (data source ECDC (9)). Circled numbers indicate epochs ① and ② of the pandemic.

Using this transition function, students estimate the number of lives that were lost because the rest of America didn't follow New York City's lead with mask wearing and social distancing. The estimate they obtain is 60,000+ lives lost by Memorial Day (May 25, 2020) based on the observed crude mortality ratio of $m_c = 0.0595$ in the ECDC data (9). This estimate is based on a simple empirical correlation students discover between the observed mortality rate and the observed infection rate $R_i$ (3,17).

## B. The summer surge

Using similar techniques, students are also able to model the summer surge resulting in the fit to the USA data up to Labor Day (September 7, 2020) that's shown in Fig 9. Students add parameters for epoch ③ with an infection rate constant $k_3$ for the relaxed social distancing at the beginning of the summer surge, a transition time $t_{23}$, and standard deviation $\sigma_{23}$. The decline in the summer surge is modeled with an infection rate constant $k_4$ for the stricter social distancing during the decline in the summer surge and a corresponding transition time $t_{34}$ and standard deviation $\sigma_{34}$.

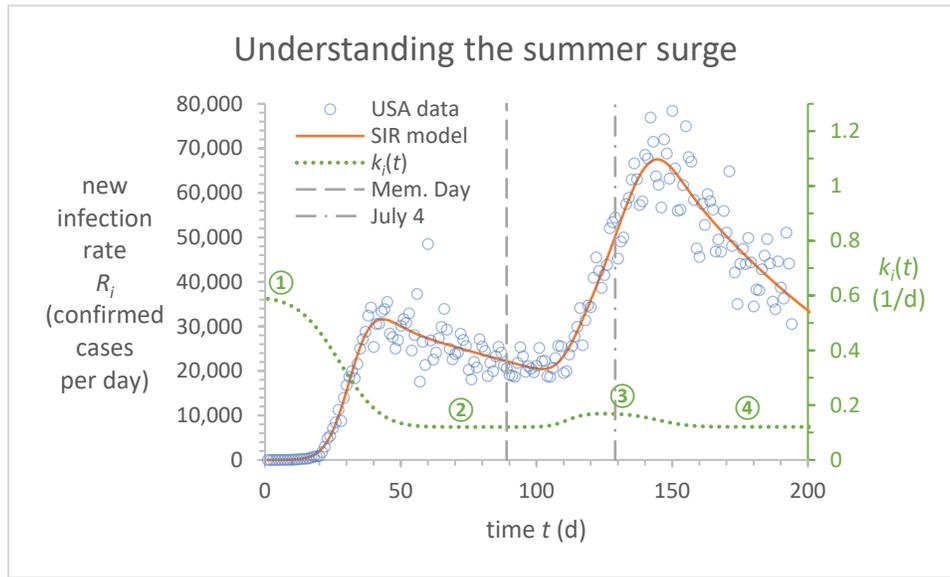

**Fig 9** Excel chart showing the predictions of the SIR model (solid orange line) when fitted to USA data reported as confirmed cases per day by the ECDC (blue circles) up to Labor Day (September 7, 2020) for four epochs of the pandemic (circled numbers) – ① the initial exponential growth, ② the epoch of social distancing, ③ the relaxation of social distancing following Memorial Day, and ④ the return to social distancing following July 4th. The vertical dashed lines indicate Memorial Day and July 4th. The graph also includes the infection rate coefficient $k_i(t)$ (green dotted line) on the secondary vertical axis (data source ECDC (9)).

## C. The fall surge and the effect of population size on the SIR model

According to a CDC report dated January 19, 2021, only 1 in 4.6 (95% UI 4.0 – 5.4) of total COVID-19 infections were reported in the period from February–December 2020 (18,19). Rounding up to 1 significant figure, that means that only about 1 in 5 of actual COVID-19 infections are represented in the data to which the model is fitted in Fig 10, i.e., $q \approx 20\%$, where $q$ is the fraction of the actual US population that is represented in the reported cases-per-day data, defined by

$$q \equiv \frac{N}{N^\star} \tag{34}$$

and $N^\star = 3.3 \times 10^8$ is the estimated actual population of the United States. There are two ways to account for this shortfall in reported cases-per-day data. One way is to multiply the cases-per-day by a factor of $1/q = 5$ and the other way is to reduce the model population size to $N = qN^\star$, or 20% of the actual US population. We chose the latter, because it makes the model predictions directly comparable to the reported cases-per-day data.

Fig 10 shows the predictions of the SIR model when students fit a fifth epoch with infection rate constant $k_5$, a transition time $t_{45}$ and standard deviation $\sigma_{45}$. The subsequent model then predicts a "fall exponential dragon" that depends on the model population size. Fig 10 shows the range of predictions of the fitted model for model population sizes of $q = 10\%$, $20\%$, and $40\%$ of the actual US population, i.e., $N = 3.3 \times 10^7$, $6.6 \times 10^7$, and $1.32 \times 10^8$ respectively.

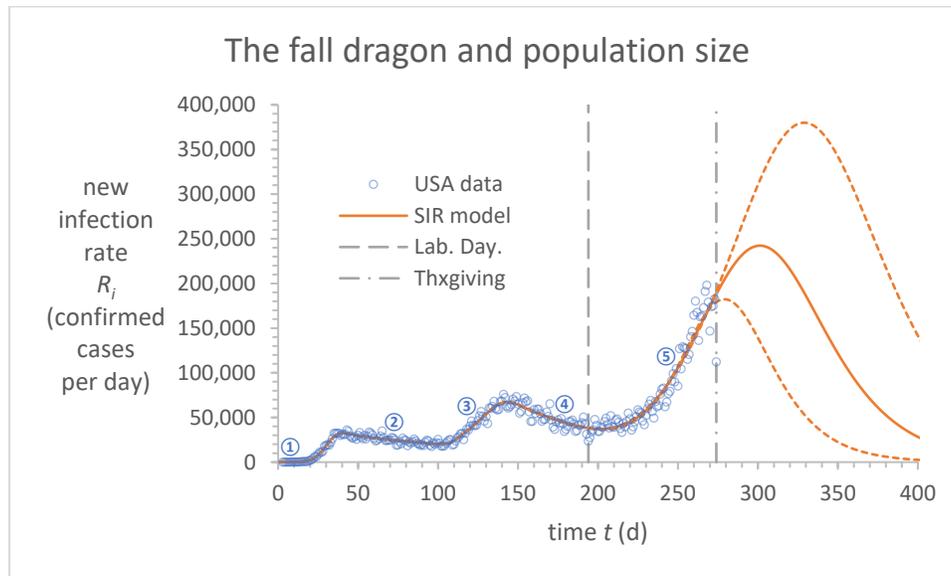

**Fig 10** Excel chart showing the predictions of the SIR model when fitted to five different epochs – ① the initial exponential growth, ② the initial period of social distancing, ③ the relaxation of social distancing following Memorial Day, ④ the return to social distancing following July 4$^{th}$, and ⑤ the fall surge following Labor Day (September 7, 2020). The blue circles show the USA data reported as confirmed cases per day up to Thanksgiving (November 26, 2020). The solid orange line represents a model population of 20% of the actual US population. The dashed orange lines represent predictions of the fitted model for model populations of 10% and 40% of the actual US population (data source OWID (20)).

As shown in Fig 10, changing the model population size and refitting has no visible effect on the fitted model up to Thanksgiving Day (November 26, 2020). However, the model predictions for the fall exponential dragon almost immediately diverge, depending on the size of the model population. In a guided-inquiry exercise, students discover that the fitted infection rate constants ($k_1 - k_5$) are different for the three fitted models and that the difference in the predicted behavior after Thanksgiving is primarily due to differences in the values of the susceptible fraction $s$ at Thanksgiving and partially due to differences in the fitted infection rate constants. The values of $s$ at Thanksgiving are $s = 0.61$, $0.80$ and $0.90$ and the fitted values of the infection rate constant are $k_5 = 0.23$, $0.18$ and $0.17$ d$^{-1}$ for models with $q = 10\%$, $20\%$, and $40\%$, respectively. The fitted value of $k_5 = 0.23$ d$^{-1}$ for $q = 10\%$ is clearly higher than the other two fits and students discover that $q = 10\%$ of the actual US population is just about the smallest model population size that's consistent with the data up to November 26, 2020 (Thanksgiving).

For the two higher fits in Fig 10, the infection rate constants are approximately the same ($k_5 \approx 0.18$ d$^{-1}$), so that the difference between them is primarily due to the difference in the susceptible fraction at Thanksgiving ($s = 0.80$ and $0.90$), respectively. Recall that the susceptible fraction $s$ is a monotonically decreasing function in the model and the $q = 20\%$ model starts closer to the peak value of $s_p = 0.70$ for epoch ⑤. In other words, the primary difference between the fits with $q = 20\%$, and $40\%$ is that there are more susceptible people left in the model population at Thanksgiving if $q = 40\%$ rather than $q = 20\%$.

## IV. MODELING VACCINATION

### A. The SIRV model

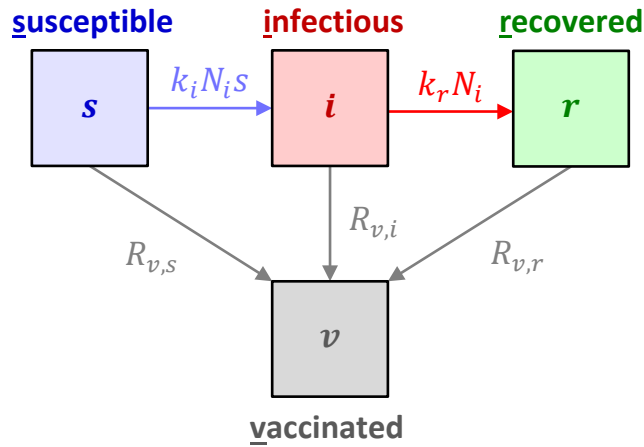

**Fig 11** FD diagram of a simple modification of the SIR model that accounts for vaccinations – the SIRV model. The four boxes represent the four parts of the model population that can be affected by the disease. Box $s$ represents the portion of the population that's susceptible to the disease. Box $i$ represents the portion of the population that's infectious. Box $r$ represents the portion of the population that has recovered from the infection (or died). Box $v$ represented the portion of the population that's been fully vaccinated.

In late December 2020, the FDA approved COVID-19 vaccines for use in the United States (emergency use authorization). Fig 11 shows a simple model of how vaccination can be added to the SIR model. The new feature is box $v$ for fully vaccinated individuals. The arrows entering box $v$ indicate the rates at which individuals are effectively vaccinated. They are then assumed to be permanently immune to COVID-19. The three rates leading to box $v$ are labeled $R_{v,s}$, $R_{v,i}$, and $R_{v,r}$ where the subscripts $s$, $i$, and $r$ indicate the originating box. These three vaccination rates are related to the to the total rate of vaccination $R_v$ in the model population by

$$R_v = R_{v,s} + R_{v,i} + R_{v,r} \tag{35}$$

The bookkeeping equation for the SIRV model is

$$N = N_s + N_i + N_r + N_v \tag{36}$$

where the subscripts once again spell out the letters of the model.

The number of vaccinated individuals in the model population is calculated from the number fully vaccinated $N_v^\star$ reported by OWID (20) using $N_v^{\text{new}} = qN_v^{\star\text{new}}$. However, because the OWID spreadsheet data contain some blank cells, the following Excel instruction is used

$$N_v^{\text{new}} = \text{IF}\left(N_v^{\star\text{new}} = 0, N_v^{\text{old}}, q * N_v^{\star\text{new}}\right) \tag{37}$$

and the FD instruction for the vaccination rate in the model population is

$$R_v^{\text{new}} = \left(N_v^{\text{new}} - N_v^{\text{old}}\right)/\delta t \tag{38}$$

The <u>r</u>ate of <u>v</u>accination of <u>s</u>usceptible individuals in the model population can be calculated using

$$R_{v,s}^{\text{new}} = N_s^{\text{old}} * R_v^{\text{new}}/\left(N_s^{\text{old}} + N_i^{\text{old}} + N_r^{\text{old}}\right) \tag{39}$$

and similarly for $R_{v,i}^{\text{new}}$, and $R_{v,r}^{\text{new}}$. Equation (39) and the corresponding equations for $R_{v,i}^{\text{new}}$ and $R_{v,r}^{\text{new}}$ assume that individuals in each of the three boxes $s$, $i$, and $r$ are equally likely to be vaccinated. Hence, in the SIRV model, the numbers in boxes, $i$ and $r$ can be calculated using

$$N_i^{\text{new}} = N_i^{\text{old}} + \left(R_i^{\text{new}} - R_r^{\text{new}} - R_{v,i}^{\text{new}}\right) * \delta t \tag{40}$$

$$N_r^{\text{new}} = N_r^{\text{old}} + \left(R_r^{\text{new}} - R_{v,r}^{\text{new}}\right) * \delta t \tag{41}$$

Combining equations (39) – (41) with bookkeeping equation (36) yields the following FD instructions for the numbers in boxes $i$, $r$ and $s$.

$$N_i^{\text{new}} = N_i^{\text{old}} + \left(R_i^{\text{new}} - R_r^{\text{new}} - N_i^{\text{old}} * R_v^{\text{new}}/\left(N - N_v^{\text{old}}\right)\right) * \delta t \tag{42}$$

$$N_r^{\text{new}} = N_r^{\text{old}} + \left(R_r^{\text{new}} - N_r^{\text{old}} * R_v^{\text{new}}/\left(N - N_v^{\text{old}}\right)\right) * \delta t \tag{43}$$

and

$$N_s^{\text{new}} = N - N_i^{\text{new}} - N_r^{\text{new}} - N_v^{\text{new}} \tag{44}$$

Using a preformatted spreadsheet, students use equations (37), (38), and (42)-(44) to implement the SIRV model and compare its predictions with reported data – see below.

**Note:** When comparing the model variables with the published data, it's important to recall that all vaccinations are reported, but only about one-in-five infections are reported.

## B. Modeling epoch ⑤ – the fall dragon

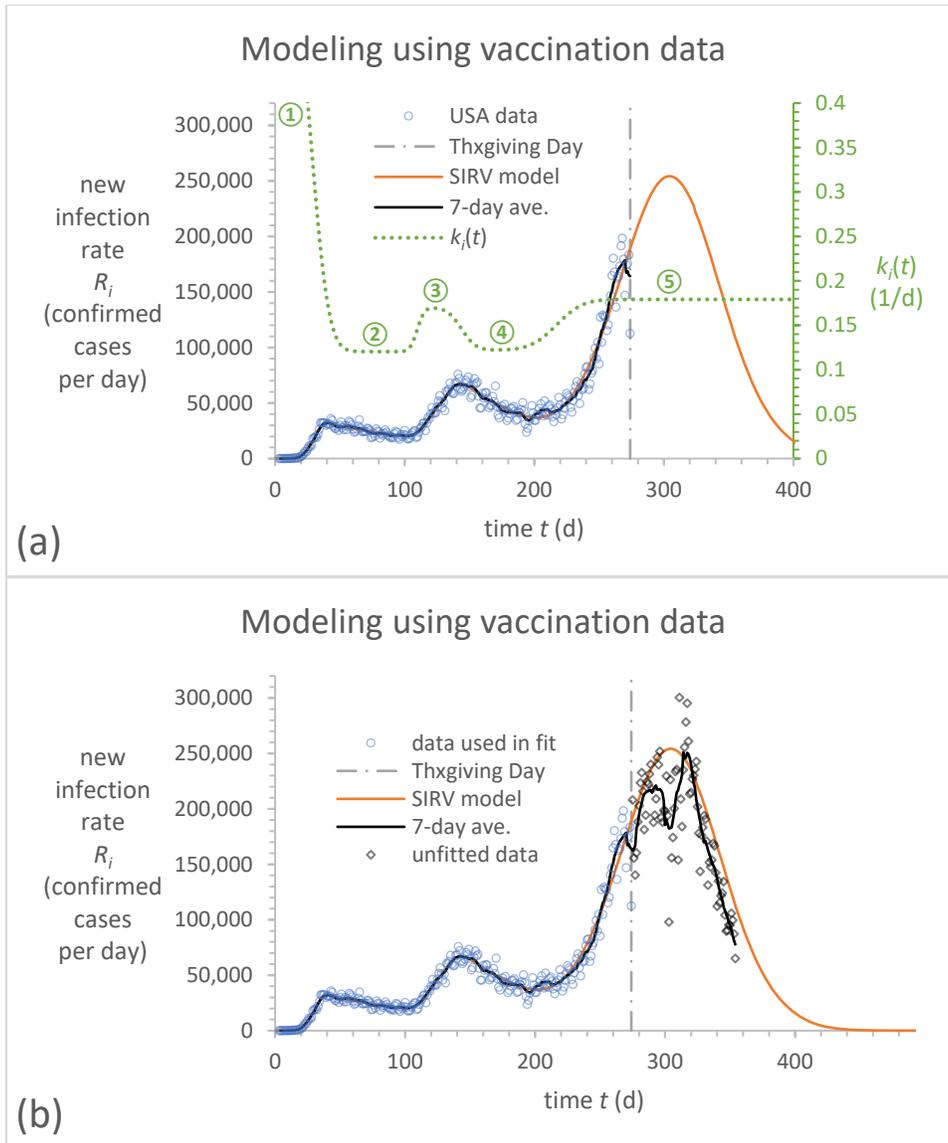

(a)

(b)

**Fig 12** Excel charts showing US data and the predictions of the SIRV model. The blue circles show USA data reported as confirmed cases per day up to Thanksgiving Day (November 26, 2020). The jagged black line shows the centered 7-day moving average of the USA data. The solid orange line shows the predictions of the SIRV model assuming that the model population is $q = 21.7\%$ of the actual population (20). Chart **(a)** shows the infection rate coefficient $k_i(t)$ as a function of time on the secondary vertical axis. Circled numbers indicate the epochs of the pandemic. Chart **(b)** shows additional USA data (grey diamonds) up to February 14, 2021, that were *not* used in the fit and the corresponding 7-day average (jagged black line). These unfitted data validate the predictions of the SIRV model with a constant infection rate coefficient of $k_5 = 0.18$ d$^{-1}$ in epoch ⑤ of the pandemic (data source OWID (20)).

Fig 12 shows the SIRV model fitted to USA data up to Thanksgiving Day (November 26, 2020) with a model population of $q = 21.7\%$ of the actual US population, i.e., $N = 7.17 \times 10^7$ based on the CDC estimate (19). The predictions of the SIRV model in Fig 12 were made using the number fully vaccinated $N_v^{\star\text{new}}$ that was reported daily by OWID (20). Fig 12(a) includes a plot of the infection rate coefficient $k_i(t)$, showing the changes in social distancing in the fitted model. An important feature of the prediction is that none of the model parameters were changed after Thanksgiving Day. Specifically, the infection rate coefficient $k_i$ remains constant at the same value $k_i = k_5 = 0.18 \text{ d}^{-1}$ that started the fall surge – throughout the entire holiday period and beyond.

Fig 12(b) shows the same fitted SIRV model as in Fig 12(a), but Fig 12(b) now includes additional USA data from Thanksgiving (November 26, 2020) through to February 14, 2021. Those additional data (grey diamonds) were *not* used in the fit and hence test the predictions of the model after Thanksgiving and throughout the 2020 holiday period and the first month and a half of 2021. Because of the large fluctuations in data reported over the holiday period, students are introduced to the idea of plotting a 7-day moving average of the USA data. As we're interested in the fit to the data, students don't use Excel's built-in moving average that averages the 7 days up to the current day (you may have seen graphs of this same type widely reported in the popular press). Instead, students use a centered moving average that doesn't produce a systematic 3-day delay in the average curve because it's centered on the current day. The centered moving average can easily be implemented using Excel's AVERAGE() function (3) and provides a good visual comparison with the model predictions. Day 400 corresponds to April 1, 2021.

## C. Epoch ⑥ and the delta variant ⑦

Fig 13 shows the SIRV model fitted to USA data up to today (August 27, 2021). Two additional epochs, ⑥ and ⑦ have been added to the model and the parameters were fitted in a similar manner to that described above. Unlike Fig 12, the fit shown in Fig 13 includes the model population size as a fitted parameter. The fitted value is $N = 6.64 \times 10^7$ or $q = 20.1\%$ of the actual US population. This number is consistent with value of $q = 21.7\%$ that was assumed for the fit in Fig 12. The LS-fit parameters for the seven epochs are recorded in Table 1. The other model parameters used in the fit were $\delta t = 1 \text{ d}$, $\tau_r = 8.0 \text{ d}$, and $N_0 = 5$. The calculated parameters corresponding to this fit are $\mathcal{R}_0 = 4.1$ and $t_d = 1.8 \text{ d}$ at the beginning of the pandemic.

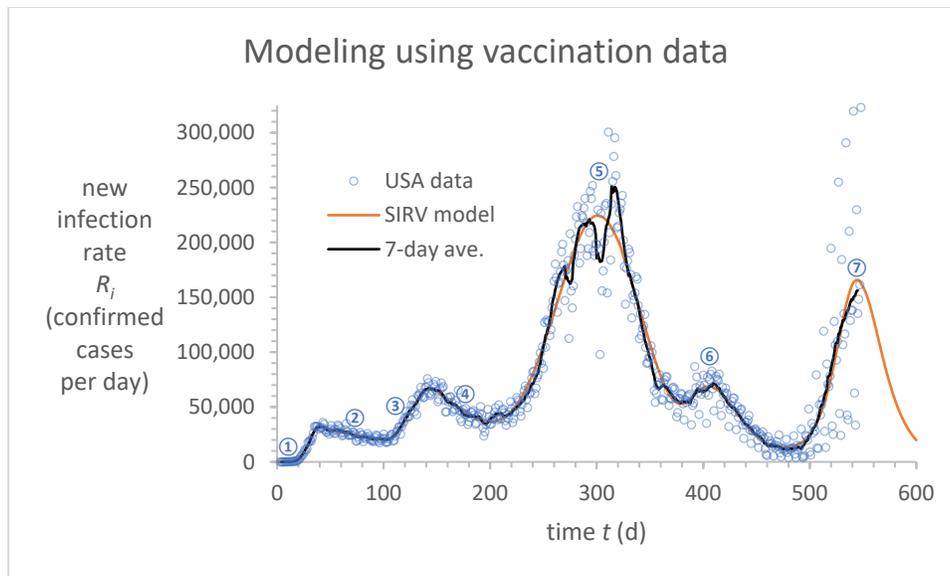

**Fig 13** Excel chart showing the predictions of the SIRV model. The blue circles show USA data reported as confirmed cases per day up to August 27, 2021. The jagged black line shows the centered 7-day moving average of the reported USA data. The smooth orange line shows the SIRV model fitted to the seven epochs of the pandemic in the US with a fitted value of $N = 6.64 \times 10^7$ or $q = 20.1\%$ of the actual US population (data source OWID (20)).

**Table 1** SIRV model parameters for the seven epochs of the COVID-19 pandemic in the US

| Epoch | start date | $t_{ij}$ (d) | $\sigma_{ij}$ (d) | $k_i$ (1/d) | $s_p = k_r/k_i$ | $h_p$ (%) |
|---|---|---|---|---|---|---|
| ① | 2/26/2020 | 0 | - | 0.59 | 0.21 | 79 |
| ② | 3/25/2020 | 28 | 11 | 0.12 | 1 | 0 |
| ③ | 6/15/2020 | 110 | 4.4 | 0.17 | 0.73 | 27 |
| ④ | 7/19/2020 | 144 | 8.8 | 0.12 | 1 | 0 |
| ⑤ | 9/29/2020 | 216 | 15 | 0.18 | 0.70 | 30 |
| ⑥ | 3/16/2021 | 384 | 10 | 0.32 | 0.39 | 61 |
| ⑦ | 7/2/2021 | 492 | 19 | 0.89 | 0.14 | 86 |

## V. DISCUSSION

"All models are wrong, but some are useful" is a common aphorism in epidemiology. In sympathy with that statement, the modeling approach presented here follows the principle of Occam's Razor, so that the models students investigate are made as simple as possible and include no unnecessary parameters. This allows our attention to be focused on the essential model assumptions and their qualitative and quantitative consequences.

According to Holmdahl and Buckee (21), COVID-19 models generally fall into one of two general categories that they call "forecasting models" and "mechanistic models". The phenomenological models discussed here are "mechanistic models" whose purpose is to gain insights into the real system being modeled. This can be contrasted with forecasting models such as the original versions of the model from the Institute for Health Metrics and Evaluation (IHME) that received a great deal of attention in the popular press in 2020 (21).

## A. Model assumptions

The models presented here can be criticized because the simplifying assumptions are clearly not 100% accurate. However, there's a long tradition of oversimplified models providing important insights into the behavior of real systems. As an example, consider the ideal gas. The assumption that gas molecules don't interact with each other is clearly wrong – gas molecules can't pass through each other, just like they don't pass through the walls of their container. However, students of thermodynamics know that the ideal gas reference state is central to the formulation of classical and statistical thermodynamics. The ideal gas model is ultimately justified *a posteriori* – by comparing its behavior with real data, and by the utility of the insights it provides. In a similar manner, the simplifying assumptions of the models presented here are ultimately justified by successful fits to the reported data, and by the utility of the insights provided by the simplified model.

**Infection is a Poisson process** All the models presented here include a transition from box $s \to i$ that's characterized by an infection rate coefficient $k_i$. In the UG model, the infection rate per infectious person is $k_i$, a constant that's independent of everything. This assumes that infectious people only interact with susceptible people and that there is an unlimited supply of susceptible people with which they can interact. In the FP model, and later SIR models, the number of interactions of infectious people with others is still assumed to be constant, but now it's assumed that the probability that the interaction is with a susceptible person, rather than another infectious person (or recovered/vaccinated individual in the SIR/SIRV models), is given by $s$ – the susceptible fraction of the entire model population. This might be a reasonable assumption for a small model population limited to a small geographic region where the entire population interacts with each other directly or indirectly on the timescale of infection and recovery, but there are clearly problems with applying this assumption to the United States as a whole. Most people stay local and don't interact with others outside of their hometown or county. Also, there are regional differences in levels of social distancing etc.

In comparing the models with reported cases-per-day data, students must interpret the meaning of the "jump" from box $s \to i$ carefully. The model jumps do not occur at the time of infection, but rather at the time that the individual becomes infectious. As with models of diffusion that include similar jumps (1), it's important for students to realize that these model jumps are not instantaneous, and they occur with a distribution of inter-event times (22).

**Recovery is a not a Poisson process** The arrows in Figs. 1, 3, 5, and 11 represent rates that depend only on the current state of the system. For example, for jumps from box $i \to r$ in the SIR-based models, the rate of recovery (or removal) is $R_r = k_r N_i$ (15), which means that the probability of any infectious person recovering is a constant – independent of everything, including how long they've been infectious (and hence been in box $i$). Because the probability is constant, the model intrinsically assumes that recovery is a Poisson process with an exponential distribution of residence times that has a peak at zero time (1,22). Students are reminded that's a good approximation for drug elimination and radioactive decay, but it's clearly not correct for recovery from COVID-19 because infectious people take a week or so to recover from the disease (even if they have mild or asymptomatic cases). However, if we consider the ensemble average number of people in box $i$ as a whole, then it seems reasonable that the ensemble average recovery rate depends directly on the number of infectious $N_i$ – if we consider times significantly longer than the mean infectious time $\tau_i$. The main advantage of equation (15) is that it's easy to understand and it gives us a simple way to predict the recovery rate $R_r$ based solely on the current number infectious $N_i$. Hence, we don't have to keep track of each individual and how long they've been infected in the model, which would be difficult to do in Excel. Students are reminded that we're only trying to understand the basics of epidemiology with our SIR model (3).

**Infection rate data** For simplicity, published cases-per-day data are compared directly with the model "infection rate" $R_i$ in these materials. This correspondence is only approximate. There is usually a delay of about a week or so between exposure to the SARS-CoV-2 virus and when they become infectious and transition to box $i$. Usually, there is a further delay before someone in the model population tests positive and appears in the published data. Hence, the time when an individual first becomes infectious (the jump from box $s \to i$) is usually somewhere between the time of exposure and the time a positive test is reported. One delay is biological – it takes time for the SARS-CoV-2 virus to be reproduced in the body to a level that makes the person infectious and that's detectable in a COVID-19 test, the other delay is behavioral/sociological – most people were not tested every day, so that the time of their positive test depends on when they were tested and when that positive test was reported and tabulated in the data used here. For example, early in the pandemic, testing resources were scarce and only symptomatic patients were tested. This uncertainty in the timing of an individual's COVID test adds to the uncertainty in the fitted transition times between epochs. Hence, the delay between exposure and becoming infectious and the delay between becoming infectious and a positive test result both need to be accounted for if one wants to interpret the correlation between public health policy changes and the fitted transition times. Further discussion of these fascinating public health policy issues is beyond the scope of this paper and is left as a research question for students.

**Gaussian transition functions** Within the model, the infection rate coefficient $k_i$ is the only model parameter that changes between epochs of the pandemic. The use of Gaussian transition functions is an empirical choice that's supported by the central limit theorem – namely that if you combine enough random events, the distribution of outcomes will approach the Gaussian (normal) distribution. However, recalling that the data being fitted is the total number of reported cases across the whole US, there are potentially many independent or correlated distributions that should be combined. Additionally, the model doesn't explicitly include the delay between exposure to the virus and becoming infectious, and the model doesn't include the subsequent delay before a positive test appears in the reported data. As a result, the fitted transition times $t_{ij}$ are likely delayed from changes in public health guidelines or mandates, and it's likely that the standard deviations $\sigma_{ij}$ of the Gaussian transition functions are too large. Both issues may be partially addressed by including a box for individuals who have been infected (exposed) but are not yet infectious – as is done in the SEIR model – see discussion below.

**Model population size** It's a bold assertion that the under-reporting of positive cases can be accounted for by a single parameter $q \equiv N/N^\star$, where $N$ is the model population size and $N^\star \approx 3.3 \times 10^8$ is the estimated actual population of the United States. The use of a single $q$ throughout the pandemic cannot be supported by direct measurement of actual infections – those data are simply not available. The best estimate I have been able to find was published by the CDC (19), but their estimates have changed over time. The most recent (July,27, 2021) estimate is 1 in 4.2 (95% UI 3.6 – 4.9) COVID–19 infections were reported from February 2020–May 2021 (19), i.e., $q = 24\%$. From a modeling perspective, the assumption that $q$ is a constant throughout the pandemic can only be justified *a posteriori* – if it's consistent with the published cases-per-day data over the entire course of the pandemic. In addition, modeling of the pandemic at the present time (late August 2021) is becoming increasing problematic as basic assumptions of the SIRV model are being shown to be incorrect – at least for some reported cases. For example, some fully vaccinated individuals are now testing positive and additional vaccination shots are planned for those previously considered to be "fully vaccinated".

The simplest way to interpret $q$ is that it's the percentage of actual infections that appear in the reported data. Using that as a measure of what's happening in the actual population assumes that those outside of the model population – spread COVID-19, are infected by COVID-19, recover from COVID-19, and are vaccinated – in a similar manner to those in the model population. This implies that individuals in the model population are mixed in with the rest of the actual population and that the rate of reported cases is proportional to the rate of actual cases.

In the SIR models, the first time that the model population size affects the qualitative behavior of the fitted model is after Thanksgiving Day (November 26, 2020) – during the fall exponential dragon (Fig 10). The reason is that the third peak in the pandemic is the first model peak that corresponds to the exponential dragon predicted by the SIR model (Fig 6(b)). The peak in the exponential dragon occurs when $s = k_r/k_i$, i.e., when the fitted model first reaches $s = s_p$ (the herd immunity threshold). For epoch ⑤, $k_i = k_5 = 0.18$ d$^{-1}$ and the peak occurs when $s = s_p = k_r/k_5 = 0.70$ or $h_p = 30\%$ (Table 1). Recall that $s \equiv N_s/N$, so that it depends on the model population size $N$. Also recall that $s$ is a monotonically decreasing function of time in all the models presented here (reinfection is not possible in all the models discussed here).

The fourth peak in the fitted SIRV model occurs in epoch ⑥ when $s = s_p = k_r/k_6 = 0.39$, or $h_p = 61\%$ (Table 1). The fact that the fit shown in Fig 13 has essentially the same value of $q$ as the fit in Fig 12, provides strong support for the SIRV model and the hypothesis that $q$ is approximately constant (at least up to the beginning of epoch ⑦). The fitted value of $k_6 = 0.32$ d$^{-1}$, is nearly double $k_3$ and $k_5$ reflecting a significant further reduction in social distancing during epoch ⑥, although the infection rate constant is still nearly half of what it was during the uncontrolled spread in epoch ① with $k_1 = 0.59$ d$^{-1}$ at the beginning of the pandemic.

The much larger fitted infection rate constant of $k_7 = 0.89$ d$^{-1}$ during epoch ⑦ can be attributed to the emergence of the delta variant of the SARS-CoV-2 virus in the United States (and the low level of social distancing). In an analogous manner to the third and fourth peaks, the model predicts that the fifth peak in the SIRV model will occur when $s = s_p = k_r/k_7 = 0.14$ or $h_p = 86\%$ (Table 1) – assuming the infection rate coefficient remains constant. The fitted infection rate constant for the delta variant ($k_i = k_7 = 0.89$ d$^{-1}$) is ~2.8 times higher than the previous variants of the virus during epoch ⑥, which, if the level of social distancing is the same, would imply that the basic reproduction number for the delta variant could be as high as $\mathcal{R}_0 \approx 11$ in the absence of any social distancing measures.

After the fitted period (up to August 27, 2021), the SIRV model makes a rather bold prediction that the curve will peak near the end of August so that the cases-per-day data are predicted to fall off during September and October 2021 (Fig 13, Day 600 corresponds to October 18, 2021). This prediction relies on the infection rate coefficient remaining constant at $k_7 = 0.89$ d$^{-1}$ and that the model population size is constant at $q \approx 20\%$ throughout the entire pandemic.

Of all the assumptions made in the SIRV model, the assumption that $q$ did not and will not change with time (and is thus a single constant throughout the pandemic) is probably the most questionable. COVID-19 tests were scarce at the beginning of the pandemic, and estimates were made that fewer than 1 in 10 cases were reported, i.e., $q < 10\%$. According, to the CDC this number increased to an average of over $q = 20\%$ by January 2021 and $q = 24\%$ from February 2020 – May 2021 in the report dated July 27, 2021 (19). Clearly, this assumption must be kept in mind when considering the model parameters particularly at the beginning and the end of the pandemic. For example, one implication of the CDC estimated increase in $q$ is that the current value of $s$ (on August 27, 2021) in the fitted SIRV model is probably too low so that the predicted delta variant peak (epoch ⑦) likely occurs too soon because there are more susceptible individuals $s$ left in the actual population than the SIRV model predicts (23).

**Vaccinations** The form of the SIRV model shown in Fig 11 was chosen to match the vaccination program in the United States. COVID-19 tests were not a prerequisite for vaccination. Hence, the status of individuals receiving vaccinations is not included in the data. As a result, the SIRV model assumes that vaccinations were administered to anyone in the population that was asymptomatic at the time. That assumption is reflected in equation (39) and the corresponding equations for $R_{v,i}^{\text{new}}$, and $R_{v,r}^{\text{new}}$, so that the rate of vaccination of individuals in each of boxes $s$, $i$, and $r$ is directly proportional to the numbers currently in each respective box. This assumption overestimates the vaccination rate of people in box $i$ (because symptomatic individuals were not supposed to be vaccinated) and underestimates vaccinations of individuals in boxes $s$ and $r$. The model does not consider partially vaccinated individuals. Recall, $N_v^{\star\text{new}}$ is the reported number of *fully vaccinated* individuals.

People are considered fully vaccinated 2 weeks after their second dose of the Pfizer-BioNTech or Moderna COVID-19 vaccines, or 2 weeks after the single-dose Johnson & Johnson's Janssen COVID-19 vaccine (24). Just like the other jumps in the SIR models, this extended process is approximated by a single jump transition of variable duration. As a result, students should once again be reminded that we're only trying to understand the basics of epidemiology with our SIRV model (3).

Finally, the initial vaccination rollout in the US was targeted at specific groups – healthcare workers, nursing home residents and the elderly, with progressively lower age restrictions until the vaccine was released for everyone 12 or older. As of late August 2021, COVID-19 vaccines have not been FDA-approved for children under 12. The models presented here don't take age into account, treating everyone in the model population in the same manner irrespective of their age. Clearly, this is another assumption that is questionable, as different age groups behaved in different ways during the pandemic and had different susceptibilities to the disease.

**Immunity is permanent?** A basic assumption of the SIR model is that recovery from COVID-19 imparts permanent immunity. There is no mechanism in the model for an individual becoming infectious a second time. That assumption is not universally correct. If immunity granted by past infection were permanent, then immunizing those who have recovered from COVID-19 would be pointless because they wouldn't gain any benefit from vaccination.

A basic assumption of the SIRV model is that being fully vaccinated always imparts permanent immunity. In late August 2021, that assumption is now known to be not accurate. There have been reported cases of COVID-19 in individuals who had been previously "fully vaccinated".

The assumption that immunity is permanent provides a basic constraint on all the SIR-based models presented here. At a constant infection rate (of any fixed value), the pandemic is predicted to peter out when enough people become immune – either by being infected (and recovering) – or by being vaccinated. According to the model, the third and fourth peaks in epochs ⑤ and ⑥ of the pandemic were caused by this affect and the predicted peak near the end of August (epoch ⑦) is also reliant on that assumption. Hence, if $k_7 = 0.89 \text{ d}^{-1}$ represents the infection rate constant for the delta variant of COVID-19 (at the current level of social distancing) – and that delta is the last variant to appear in the data – then in late August 2021 we are closing in on a final test of the permanent immunity hypothesis in our fitted SIRV model and the related assumption that $q$ is a constant throughout the pandemic.

### B. Educational objectives and scope

The teaching materials discussed here are designed as a case study in modeling a complex data set. They are not meant to directly inform public policy. However, simple models often provide useful insights into complex phenomena – not just by what they model successfully, but also by what they *cannot* explain. These insights are not usually provided by forecasting models (21).

**Finite difference methods** Students discover that the simple finite difference methods that they first learned in the context of molecular biophysics can also be applied to epidemiological models of COVID-19 in the United States. All the models that students investigate predict exponential growth at the beginning of the pandemic. Exponential growth is qualitatively different from most models in molecular biophysics. Comparison of exponential growth with exponential decay provides students with further insights into the properties of all models that predict proportional change (1), and into challenges with the accuracy of the finite difference method during rapid exponential growth (3).

**Systematic model development and least-squares fits** The teaching materials discussed here can be used as an introduction to performing least squares (LS) fits. Students are guided through the process of calculating the residuals between observed data and the predictions of the model, calculating $Q$, the sum of the squares of the residuals, and then using Excel's Solver to find the minimum in $Q$ (3). If needed, students can also be directed to Chapter 6 of ref. (1) for a more detailed introduction to least-squares fits in the context of $O_2$ binding to myoglobin.

As mentioned in the introduction to this paper, performing LS fits to complex data is more of an art than a science, particularly when the model predictions depend exponentially on the fitted parameters (5). Utilizing the principle of Occam's Razor is central to the approach presented here.

Students are guided through the modeling process starting with fitting the UG model to epoch ① with Excel's "Exponential Trendline" feature and then an LS fit. They then fit the SIR model to epoch ② (after the transition is complete) to reproduce Fig 7. In producing the LS fit shown in Fig 8, students first estimate $t_{12}$ and $\sigma_{12}$ "by hand" – adjusting their values in the spreadsheet and observing the effect on the model predictions. Only once they have an approximate fit, do students use Excel's Solver to find the minimum in $Q$. Once students have a fit as shown in Fig 8, they systematically investigate how changing model parameters $N$ and $\tau_i$ affects the model and it's fitted parameters. They discover that the model, up to the end of epoch ②, can be successfully fitted with any reasonable values of $N$ and $\tau_i$. The remainder of the fits to the SIR and SIRV models are done in a similar manner by systematically adding one epoch at a time. That approach ensures that students understand how each added parameter affects the model and helps them avoid lack-of-convergence problems that can plague complex LS fits.

Least-squares fits using the SIR model are not like arbitrary polynomial fits. Polynomials can be fitted to almost any shape curve, but the SIR model (with constant $k_i$) always predicts a characteristic exponential dragon shape for the infection rate $R_i(t)$ – see Fig 6(b). After a transition to a new epoch, the shape of the $R_i(t)$ curve is determined by the current value of the susceptible fraction $s$ (and the infectious fraction $i$) and the new value of the infection rate coefficient $k_i$ because all the other model parameters are held constant throughout the pandemic. As students discover, epoch ① corresponds to exponential growth at the beginning of the dragon; epoch ② corresponds to a gradual exponential decay during the dragon's tail caused by $s < k_r/k_2$, $(h > h_p)$ with low $k_i$; epoch ③ corresponds to exponential growth with $s > k_r/k_3$, $(h < h_p)$; epoch ④, like epoch ②, corresponds to a gradual exponential decay caused by $s < k_r/k_4$, $(h > h_p)$ with low $k_i$; epoch ⑤ corresponds to an exponential dragon where the susceptible fraction starts with $s > k_r/k_5$, $(h < h_p)$ and transitions to $s < k_r/k_5$, $(h > h_p)$ as the susceptible fraction decreases and passes through $s = k_r/k_5 = s_p$ (herd immunity for $k_i = k_5$); epoch ⑥ corresponds to an exponential dragon in which the susceptible fraction decreases and passes through the value of $s = k_r/k_6 = s_p$ (herd immunity for $k_i = k_6$); finally, epoch ⑦ (fitted up to August 27, 2021), corresponds to another exponential dragon that's predicted to peak around the present time (August 27, 2021) when the susceptible fraction passes through the value of $s = k_r/k_6 = s_p$ (herd immunity for $k_i = k_7$) – Fig 13 and Table 1.

The fit to epoch ① is a foundational confirmation that the UG model (and the subsequent SIR models) are reasonable as they successfully predict exponential growth at the beginning of the pandemic. The fits to epochs ②, ③ and ④, are not particularly impressive from a modeling perspective – as the fitted model does not appear significantly different from a straight line (outside of the Gaussian transition periods). Almost any model can predict linear behavior.

However, the fact that the fitted model predicts a peak in epoch ⑤ that has the correct approximate timing, height, and width – with a single constant value of $k_i = k_5 = 0.18 \text{ d}^{-1}$ – is a strong validation of the SIR model during that time. Recall that the SIR model always predicts an exponential dragon peak for $R_i(t)$ of the form shown in Fig 6(b) (if $k_i$ is constant). Similarly, the fact that the SIRV model successfully models the peak in epoch ⑥ with a fitted value of $q \approx 20\%$ (in agreement with Fig 12) and a constant infection rate constant $k_6$ through the peak is another strong validation of the SIRV model. Finally, the SIRV model also successfully explains the emergence of the delta variant in epoch ⑦ and makes a rather bold prediction that the peak in $R_i(t)$ will be reached near late August 2021, assuming the infection rate coefficient doesn't increase beyond the fitted value of $k_7 = 0.89 \text{ d}^{-1}$ and that the fraction of the actual US population that's represented in the reported cases-per-day data remains constant at $q \approx 20\%$.

The model prediction that the delta variant peak is upon us at the end of August 2021 and that the cases-per-day data should fall off rapidly during September and October 2021 is based on the increasing questionable assumption that the model population is constant at $q \approx 20\%$ throughout the pandemic (and that the infection rate coefficient will remain constant at $k_i = k_7 = 0.89 \text{ d}^{-1}$). As we've already discussed, the CDC report dated July 27, 2021, estimates that $q$ is increasing as a larger fraction of actual cases are reported (19). Hence, it's likely that the height of the predicted delta-variant peak occurs too soon because there are more susceptible individuals left in the actual population than a value of $q \approx 20\%$ predicts. In addition, it's likely that the value of $k_7 = 0.89 \text{ d}^{-1}$ is an overestimate in a similar manner to the fitted value of $k_5 = 0.23 \text{ d}^{-1}$ being too high in the fit shown in Fig 10 with $q = 10\%$. The validity of the assumption that $q \approx 20\%$ throughout the pandemic will be tested in the next few weeks or so (3).

In summary, the systematic least-squares approach enables students to appreciate that the SIR model and it's SIRV variant do a surprisingly good job of modeling the pandemic in the United States from February 26, 2020, through August 27, 2021. The most important question is not what's wrong with the over-simplified model, but rather – why does it work so well?

## C. Model extension – the SEIR model

An obvious extension to the work presented here is to change the base SIR model to the SEIR model. The main new feature of the SEIR model is the explicit inclusion of a box $e$ for <u>e</u>xposed individuals who have been infected but are not yet infectious. SEIR model box $e$ is inserted between boxes $s$ and $i$ of the original SIR model. Modeling using the SEIR model is not included in this case study because it would add another adjustable parameter to the base model and following the principle of Occam's Razor it's been omitted. Further investigation of the SEIR model and it's SEIRV variant is left as a research exercise for students. However, a preliminary investigation has shown that the SEIR model also needs a Gaussian transition function to successfully model the transition from epoch ① to epoch ② ...

## VI. CONCLUSION

This project was motivated by a question. Can undergraduates use spreadsheets to successfully model the spread of COVID-19? The answer is a resounding "yes!" In fact, this topic makes an excellent capstone experience for students interested in scientific modeling. The finite difference (FD) methods used are accessible to students at the level of introductory physics and they reinforce the universal applicability of computational methods in scientific modeling. Along the way, students gain a different perspective on kinetic models and rate constants by applying them to the behavior of people. While people don't jiggle around like molecules in solution, they do have interactions with others at a rate that can be successfully modeled using familiar biophysical techniques.

Excel is an often-underrated platform for computational modeling. It has numerous advantages for undergraduate students and their instructors that facilitate the learning objectives of this case study. Excel is familiar and non-threatening to students – most undergraduates have already used it to plot data in their science labs. It also has many features that make it ideal for modeling the spread of COVID-19. The most obvious feature is that calculations are laid out spatially, which makes it easier for students without programming experience to follow the logic of the computational approach. Another advantage is the ease of graphing.

FD methods can be easily implemented in spreadsheets (1), allowing students to understand and calculate solutions to differential models that have no analytical solution. In addition, there's an extremely simple procedure for performing least-squares fits to computational models using Excel's Solver (1,25). Hence, Excel is an excellent platform for practical reasons and because it lets students, and their instructors, focus on the learning objectives of – 1. Developing FD methods; 2. Using systematic model development techniques; and 3. Validating the models by fitting them to real data using least-squares. A scaffolded guided-inquiry approach is used so that students are actively engaged in investigating the consequences of the model assumptions in a systematic step-by-step manner. That approach facilitates student understanding of the FD models as they develop them, and it enables students to see how the model parameters affect the qualitative and quantitative predictions of these introductory models as they are systematically developing them, while simultaneously validating them using least-squares fits to reported data.

Even though the approach uses only introductory methods, the modeling approach is surprisingly successful in modeling the spread of COVID-19 in the US. Because of its simplicity, the model also provides unexpected insights into the spread of the virus. Notably, (after the initial exponential outbreak) the behavior of the US population up to February 14, 2021, can be separated into two categories – "stricter social distancing" and "relaxed social distancing". Epochs ② and ④ of the model in Fig 12(a) correspond to stricter social distancing with $k_2 \approx k_4 = 0.122 \pm 0.001$ d$^{-1}$ and epochs ③ and ⑤ of the model correspond to more relaxed social distancing with $k_3 \approx k_5 = 0.176 \pm 0.006$ d$^{-1}$. Hence, the inception of both the summer and fall surges can be explained by a modest 30% increase in the infection rate coefficient $k_i$.

A significant feature of the modeling approach is that it uses as few parameters as possible to model the published data. It's easy for students (and instructors) to be seduced into the notion that the model can do better – and it can – but every time an additional parameter is added, the question that should be asked is – will we learn anything new from it (5,27)? As the data came in day-by-day it seemed clear to me that there were changes in the infection rate coefficient occurring during the beginning of the fall surge. However, it turned out that on a longer timescale, a single transition function could fit the data almost as well, and with a simpler and more insightful observation that $k_5 \approx k_3$.

As the model is extended beyond Thanksgiving (November 26, 2020), students discover that the size of the model population $N$ becomes an important parameter in the fitted model. As shown in Fig 12(b), the projected model, with a model population of $q \approx 20\%$ of the actual US population, appears to match the USA data quite well. Once vaccinations began, students added vaccination to the SIR model resulting in an SIRV model that explicitly includes a separate box $v$ for the fully vaccinated. The success of the models in predicting the basic shape, height and timing of the third peak (Fig 12) is a significant validation of the predictions of the SIR model and it's SIRV variant because they have no wiggle room in the form of the predicted exponential dragon assuming constant $k_i = k_5$ and a constant value of $q \approx 20\%$.

The success of the SIRV model in explaining the fourth smaller peak in the pandemic (Fig 13) with $k_i = k_6$ and a fitted value of $q \approx 20\%$ (the same as the rest of the pandemic), is a significant additional validation of the SIRV model. A final test of the SIRV model is whether it can explain the increase in the cases-per-day data during the beginning of epoch ⑦. As shown in Fig 13, the SIRV model can not only fit the data with the same value of $q \approx 20\%$ but it also provides insights into just how infectious the delta variant is compared with the original variants of the SARS-CoV-2 virus. Only time will tell if the SIRV model's predictions for the head and tail of the delta-variant exponential dragon are correct.

While the five peaks in Fig 13 have a similar appearance, it's important to note that the third, fourth and fifth peaks in Fig 13 are qualitatively different from the first two peaks. The first peak is caused by the transition from uncontrolled spread ① to the first period of stricter social distancing (epoch ②). The second peak is similarly caused by a transition from relaxed social distancing (epoch ③) to a second epoch ④ of stricter social distancing. In contrast, the third, fourth and fifth peaks, during the middle of epochs ⑤, ⑥ and ⑦ of Fig 13, are simply exponential dragons (Fig 6(b)) that are intrinsic to the SIR model with a constant infection rate coefficient. The fitted infection rate constant in epoch ⑥ is nearly twice that of the earlier epochs ③ and ⑤, indicating a further substantial reduction in social distancing measures.

The fitted infection rate constant in epoch ⑦ is ~2.8 times higher than epoch ⑥, consistent with the delta variant being ~2.8 times more transmissible than the original variants of the SARS-CoV-2 virus. The SIRV model makes a bold prediction that the peak caused by the delta variant is upon us at the end of August 2021 and that the cases-per-day data should fall off rapidly during September and October 2021. Those predictions are based on the increasingly questionable assumption that the model population is constant at $q \approx 20\%$ throughout the pandemic (and that the infection rate coefficient will remain constant at $k_i = k_7 = 0.89 \text{ d}^{-1}$). Only time will tell if those assumptions remain applicable.

The success of the SIRV model in explaining and predicting the quantitative behavior of the spread of COVID-19 from February 26, 2021, through August 27, 2021, is a significant validation of the basic SIR model and its SIRV variant. As students discover, people aren't molecules … but sometimes they behave like them.

## ACKNOWLEDGMENTS

I wish to thank Paul Ginsparg, Robert Hilborn, and Jaqueline Lynch for helpful comments on earlier drafts of the manuscript. Support from the National Institutes of Health (Fellowship GM20584) and the National Science Foundation (Grant Nos. 0836833 and 1817282) is gratefully acknowledged.

## APPENDIX – SYMBOL GLOSSARY

**Table A.1** Non-letter symbols

| Symbol | Description |
|---|---|
| [=] | a symbol that's pronounced "has units of" |
| ≈ | alternate equals sign that means "is approximately" |
| ≡ | alternate equals sign that means "is defined as" |

**Table A.2** Prefixes, suffixes, subscripts, and superscripts

| Symbol | Description |
|---|---|
| $(t)$ | suffix used to indicate a function of time, e.g., $R_i(t)$. |
| 0 | subscript "naught" meaning zero. The value of a variable at time zero, e.g., $N_0$. |
| $\delta$ | prefix used to indicate a small change in a finite difference (FD) algorithm. |
| new | superscript used to indicate the current step in an algorithm, which corresponds to the current row in a spreadsheet |
| old | superscript used to indicate the previous step in an algorithm, which corresponds to the previous row in a spreadsheet |

**Table A.3** Letter and letter-like symbols

| Symbol | Description |
|---|---|
| $\sigma_{12}, \sigma_{23}, \sigma_{34}, \ldots \,[=]\, \text{d}$ | $\sigma_{12}$ is the **s**tandard deviation of the transition time between epochs 1 and 2, similarly for $\sigma_{23}$ etc. |
| $\tau_i \,[=]\, \text{d}$ | mean **i**nfectious **t**ime – the average time a person is infectious in the SIR model |
| $F_{12}, F_{23}, F_{34}, \ldots \,[=]\, 1$ | $F_{12}$ is the cumulative probability of the Gaussian transition **f**unction between epochs 1 and 2, similarly for $F_{23}$ etc. |
| $h = 1 - s \,[=]\, 1$ | fraction immune – the fraction of the model population that's immune from infection |
| $h_p = 1 - s_p \,[=]\, 1$ | herd immunity threshold – the fraction immune required for decline in the number infectious |
| $i \equiv \dfrac{N_i}{N} \,[=]\, 1$ | fraction **i**nfectious – the fraction of the model population that's infectious |
| $k_1, k_2, k_3, \ldots \,[=]\, \dfrac{1}{\text{d}}$ | infection rate constants for the COVID-19 SIR and SIRV models in epochs 1, 2, 3 … |
| $k_i \,[=]\, \dfrac{1}{\text{d}}$ | **i**nfection rate constant (or coefficient) for all of the COVID-19 models |
| $k_r \,[=]\, \dfrac{1}{\text{d}}$ | **r**ecovery rate constant for the SIR model |
| $N \,[=]\, 1$ | total **n**umber of people in the model population |
| $N^\star \,[=]\, 1$ | total **n**umber of people in the actual United States population |
| $N_i \,[=]\, 1$ | **n**umber **i**nfectious in the model population |
| $N_r \,[=]\, 1$ | **n**umber **r**ecovered in the model population |
| $N_s \; [=] \; 1$ | <u>n</u>umber <u>s</u>usceptible in the model population |
| $N_v \; [=] \; 1$ | <u>n</u>umber <u>v</u>accinated in the model population |
| $N_v^\star \; [=] \; 1$ | <u>n</u>umber <u>v</u>accinated in the actual United States population |
| $p_{12}, p_{23}, p_{34}, \ldots \; [=] \; \frac{1}{d}$ | $p_{12}$ is the probability density of the Gaussian transition <u>f</u>unction between epochs 1 and 2, similarly for $p_{23}$ etc. |
| $q \equiv \frac{N}{N^\star} \; [=] \; 1$ | fraction of the US population included in the model population (fraction of cases per day that are reported) |
| $\mathcal{R}_0 \equiv k_i \tau_i = \frac{k_i}{k_r} \; [=] \; 1$ | basic <u>r</u>eproduction number of the SIR model – the average number of people infected by an infectious individual in a completely susceptible population |
| $R_i \; [=] \; \frac{1}{d}$ | the <u>i</u>nfection <u>r</u>ate (reported cases per day) |
| $R_r \; [=] \; \frac{1}{d}$ | the <u>r</u>ecovery <u>r</u>ate of infected individuals in the model population |
| $R_v \; [=] \; \frac{1}{d}$ | the effective <u>v</u>accination <u>r</u>ate of individuals in the model population |
| $R_{v,s} \; [=] \; \frac{1}{d}$ | the effective <u>v</u>accination <u>r</u>ate of individuals in box <u>s</u> of the model population, similarly for $R_{v,i}$ and $R_{v,r}$ |
| $s \equiv \frac{N_s}{N} \; [=] \; 1$ | fraction <u>s</u>usceptible – the fraction of the model population that's still susceptible to infection |
| $t \; [=] \; d$ | time variable in the epidemiological models that starts at $t = 0$ |
| $t_d \; [=] \; d$ | <u>d</u>oubling <u>t</u>ime – the time it takes for the number infectious $N_i$ to double |

## REFERENCES

1. Nelson, Peter Hugo. 2021. Biophysics and Physiological Modeling. http://circle4.com/biophysics/
2. In principle, any spreadsheet compatible with Microsoft Excel could be used. However, neither my students nor I have tested the methodology on other platforms such as Google Sheets.
3. Nelson, Peter Hugo. 2021. June 20, 2021. Biophysics and Physiological Modeling Chapter 12: COVID-19 and epidemiology http://www.circle4.com/biophysics/chapters/BioPhysCh12.pdf
4. Nelson, Peter Hugo. 2012. Teaching introductory STEM with the Marble Game, arXiv, arXiv:1210.3641, https://arxiv.org/abs/1210.3641 (preprint posted October 22, 2012)
5. Nelson, Peter Hugo. 2011. A permeation theory for single-file ion channels: One-and two-step models. J. Chem. Phys. 134:165102.
6. I chose to use the symbol $N_i$ for the number infectious to make the notation match standard biochemical practice for molecular systems. However, epidemiologists prefer using the single uppercase letter $I$ instead of $N_i$. They also prefer to use Greek letters for the rate constants so that the infection rate is written as $\beta I$ (7). We'll stick with using the traditional chemistry $k$ with a descriptive subscript for rate constants and $N$ with a descriptive subscript for numbers in the boxes of the models.

7.  Jones, James Holland. 2007. February 20, 2021. Notes On $\mathcal{R}_0$ https://web.stanford.edu/~jhj1/teachingdocs/Jones-on-R0.pdf
8.  Nelson, Peter Hugo. 2020. February 20, 2021. 12.1 The coronavirus outbreak - exponential growth https://youtu.be/gLao39Wcf3Y
9.  European Centre for Disease Prevention and Control (ECDC). 2020. December 15, 2020. European Centre for Disease Prevention and Control – Download today's data on the geographic distribution of COVID-19 cases worldwide https://www.ecdc.europa.eu/en/publications-data/download-todays-data-geographic-distribution-covid-19-cases-worldwide
10. Mathematical Association of America. 2021. April 7, 2021. https://www.maa.org/book/export/html/115630
11. Kermack, William Ogilvy, and Anderson G. McKendrick. 1927. A contribution to the mathematical theory of epidemics. Proceedings of the Royal Society of London. Series A, Containing papers of a mathematical and physical character. 115(772):700-721.
12. The idea of using a dragon analogy for explosive exponential growth was inspired by the expression "tickling the dragon's tail" that's based on a remark by Richard Feynman about the dangers of some ill-advised early nuclear experiments – where exponential growth had the potential for similar catastrophic consequences. See section 12.4 of ref. (3).
13. Metcalf, C. Jessica E., M. Ferrari, Andrea L. Graham, and Bryan T. Grenfell. 2015. Understanding herd immunity. Trends in Immunology, 36(12):753-755.
14. Heesterbeek, Johan Andre Peter. 2002. A brief history of R_0 and a recipe for its calculation. Acta Biotheoretica 50(3):189-204.
15. Hilborn, Robert C. 2020. Personal communication.
16. This is the simplest assumption for the distribution of transition times. Other distributions could be chosen, or external data sources could be used to inform the choice.
17. Nelson, Peter Hugo. 2020. February 20, 2021. 12.6 Lives lost because people didn't wear masks. https://youtu.be/iTFnGjsnlgg
18. Heather Reese, A. Danielle Iuliano, Neha N. Patel, Shikha Garg, Lindsay Kim, Benjamin J. Silk, Aron J. Hall, Alicia Fry, Carrie Reed. 2020. Estimated Incidence of Coronavirus Disease 2019 (COVID-19) Illness and Hospitalization—United States, February–September 2020, Clinical Infectious Diseases, 2020:ciaa1780. https://doi.org/10.1093/cid/ciaa1780
19. Centers for Disease Control and Prevention (CDC). 2021. April 7, 2021 Estimated Disease Burden of COVID-19 https://www.cdc.gov/coronavirus/2019-ncov/cases-updates/burden.html, https://web.archive.org/web/20210401182614/https://www.cdc.gov/coronavirus/2019-ncov/cases-updates/burden.html
20. Our World in Data (OWID). 2021. August 27, 2021. United States: Coronavirus Pandemic Country Profile https://ourworldindata.org/coronavirus/country/united-states
21. Holmdahl, Inga, and Caroline Buckee. 2020. Wrong but Useful — What Covid-19 Epidemiologic Models Can and Cannot Tell Us. N. Engl. J. Med. 383(4):303-305.
22. Nelson, Peter Hugo, Alan B. Kaiser, and David M. Bibby. 1991. Simulation of diffusion and adsorption in zeolites. J. Catal. 127(1):101-112.
23. Fig 10 illustrates this same delay and increase in the size of the predicted exponential dragon as $q$ is increased from $q = 10\%$ to $q = 20\%$. In addition, the fitted value of $k_7$ is probably too high, in a similar manner to the fitted value of $k_5 = 0.23 \text{ d}^{-1}$ being too high in the fit with $q = 10\%$ (Fig. 10).

24. Centers for Disease Control and Prevention (CDC). 2021. August 19, 2021. Key Things to Know About COVID-19 Vaccines https://www.cdc.gov/coronavirus/2019-ncov/vaccines/keythingstoknow.html
25. Although the approach is aimed at students without formal programming experience, it can be easily adapted for students with programming languages such as Python (26).
26. Ginsparg, Paul. 2021. June 11, 2021. Info 3950 Problem Set "9" https://nbviewer.jupyter.org/url/courses.cit.cornell.edu/info3950_2021sp/ps9.ipynb
27. Nelson, Philip Charles. 2004. Biological Physics, Energy, Information, Life. W. H. Freeman and Company, New York.